\documentclass[12pt,english]{article}
\usepackage{lmodern}
\usepackage[T1]{fontenc}
\usepackage[latin9]{inputenc}
\usepackage{amsmath}
\usepackage{amssymb}
\usepackage{esint}

\usepackage{geometry}
\makeatletter
\usepackage{float} 
\usepackage[colorlinks=true, linkcolor=ForestGreen, citecolor=blue]{hyperref}
\usepackage{cite}
\usepackage{babel}
\usepackage{mathdots}
\usepackage{amsfonts}
\usepackage{mathrsfs}
\usepackage{amsmath,booktabs}
\usepackage{esint}
\usepackage{graphicx}
\usepackage{youngtab}
\usepackage{multicol}
\usepackage{multirow}
\usepackage{slashed}
\usepackage{simplewick}
\usepackage{braket}
\usepackage{bm}
\usepackage{relsize}
\usepackage[dvipsnames,table,xcdraw]{xcolor}
\usepackage{subfigure}
\usepackage{feynmp}
\usepackage{pifont}
\usepackage{cancel}
\numberwithin{equation}{section}
\allowdisplaybreaks

\makeatother

\usepackage[dvipsnames]{xcolor}

\begin{document}
\author{Wan-Zhe Feng\footnote{Email: vicf@tju.edu.cn}~~and~Zi-Hui Zhang\footnote{Email: zhangzh\_@tju.edu.cn}\\
\textit{\small{Center for Joint Quantum Studies and Department of Physics,}}\\
\textit{\small{School of Science, Tianjin University, Tianjin 300350, PR. China}}}
\title{Annihilating to the Darker: Thermal Relic Dark Matter with an Ultraweak Portal to the Standard Model}
\date{}

\maketitle

\begin{abstract}

Thermal relic dark matter has been severely constrained in recent years by direct and indirect dark matter searches, as well as multi-messenger probes of dark sectors. At the current level of experimental precision, it has become difficult for many thermal dark matter models to deplete their abundance sufficiently through freeze-out to reproduce the observed relic density. We study the possibility that thermal dark matter couples only ultraweakly to the Standard Model (SM), and therefore remains effectively undetectable in current experiments, while interacting much more strongly with a darker sector that controls its freeze-out history. Hence, the dominant annihilation channels of a thermal relic may proceed primarily into the darker sector rather than into SM particles. We first summarize the general classes of portal interactions that may connect the SM, a hidden sector, and a darker concealed sector, together with the corresponding experimental constraints. We then illustrate the mechanism in two representative realizations. The first is a prototype $U(1)_x \times U(1)_c$ setup with kinetic and mass mixing between the hidden and concealed gauge sectors. The second is a more motivated $U(1)_{B-L}\times U(1)_c$ construction, in which the $U(1)_{B-L}$ gauge interaction is strongly constrained and the hidden--concealed connection is mediated primarily by a real scalar. In both frameworks, we identify two qualitatively distinct scenarios: \emph{assisted depletion}, in which the hidden sector dark matter remains the dominant component but its abundance is efficiently reduced with the assistance of the darker sector; and \emph{darker conversion}, in which the hidden sector abundance is transferred predominantly into a more secluded dark relic. By solving the full set of coupled Boltzmann equations and presenting benchmark models for dark matter masses in the 1--200~GeV range, we show that electroweak scale thermal relic dark matter may remain viable even when its direct portal to the SM is ultraweak, provided that sufficiently strong hidden--concealed interactions govern the cosmological evolution.

\end{abstract}

\newpage{}

{  \hrule height 0.4mm \hypersetup{colorlinks=black,linktocpage=true}
\tableofcontents
\vspace{0.5cm}
 \hrule height 0.4mm}

\newpage{}

\section{Introduction}\label{sec:Intro}

The null detection of weakly interacting massive particles (WIMPs) as dark matter candidates
has posed a serious challenge to a wide range of dark matter models.
In particular, the steadily increasing sensitivity and precision of
direct detection experiments~\cite{XENON:2019zpr,XENON:2019gfn,DarkSide-50:2022qzh,PandaX-4T:2021bab,LZ:2022lsv,XENON:2023cxc,LZ:2024zvo}
and indirect searches~\cite{Planck:2015fie,AMS:2021nhj,McDaniel:2023bju,CTA:2020qlo}
have significantly narrowed the viable parameter space of most conventional WIMP scenarios.
The reason is remarkably straightforward:
to obtain an annihilation cross section large enough to deplete the dark matter abundance to the observed relic density,
the couplings between dark matter and Standard Model (SM) particles must typically be sizable.
Such couplings in turn lead to substantial elastic scattering cross sections between dark matter and hadrons,
many of which have already been excluded by direct detection experiments.\footnote{
An exception arises in quark-phobic dark matter models, in which dark matter does not annihilate into quarks
and is therefore subject to much weaker direct detection constraints.
In the present work, we focus on dark matter models that are strongly constrained by direct detection experiments,
together with indirect searches and other experimental probes.}
Additionally, the same large dark matter annihilation cross sections also produce noticeable indirect detection signals,
which have not yet been observed.

Dark matter may originate from one or more hidden sectors,
which arise naturally in a wide class of grand unified theories and string-motivated constructions.
Since dark matter has so far been observed only through its gravitational effects,
a natural possibility is that dark matter does not communicate primarily with the SM,
but rather with other darker states.\footnote{Previous research on multiple hidden sectors within the freeze-in scenario includes~\cite{Aboubrahim:2021ycj,Aboubrahim:2022bzk,Bhattiprolu:2023akk,Feng:2024nkh,Bhattiprolu:2024dmh}.}
In particular, the interaction that determines the visibility of dark matter need not be the same as the interaction that determines its freeze-out history.
The portal interactions connecting the dark sector to the SM are strongly constrained by experimental data and are thus ultraweak,
whereas the interactions connecting the dark sector to a darker sector are much less constrained and may remain large enough to dominate the early Universe dynamics.

\begin{figure}[h!]
\begin{center}
\includegraphics[scale=0.36]{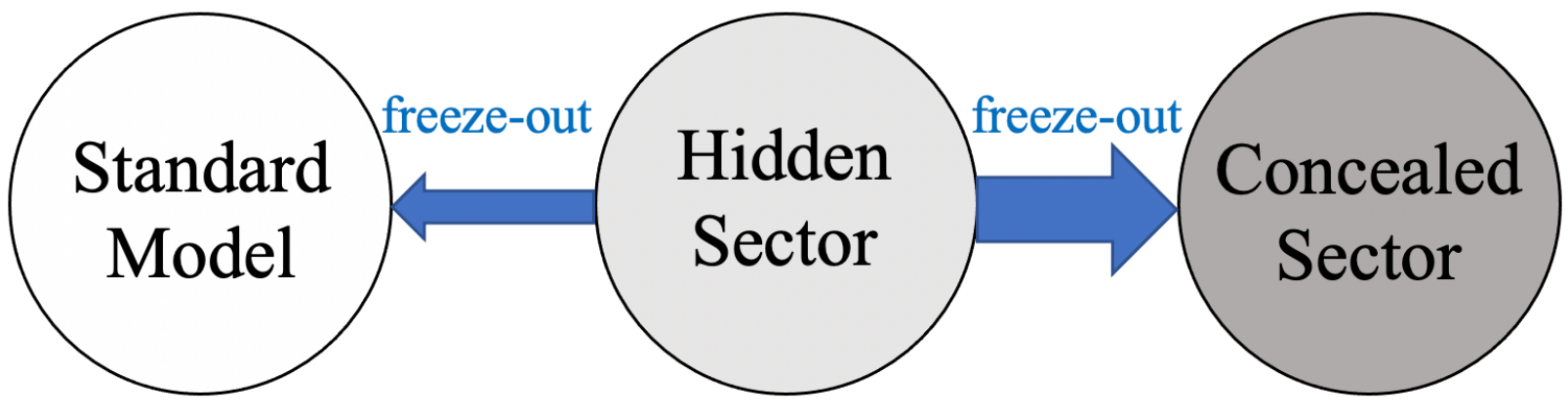}
\caption{
  \emph{Annihilating to the darker:}
  Dark matter in the hidden sector predominantly freezes out into a darker concealed sector rather than into the SM.}
\label{Fig:Model}
\end{center}
\end{figure}

In this work, we focus on dark matter candidates with masses in the range 1--200~GeV,
a region tightly constrained by current dark matter searches.
Dark matter models with candidates in this mass range are also constrained by complementary indirect searches
and by other experimental probes of the dark sector, including searches for dark mediators.
We consider a dark matter candidate residing in a hidden sector that is connected to the SM through an ultraweak portal,
but interacts more strongly with a darker concealed sector, as illustrated in Fig.~\ref{Fig:Model}.
This setup represents a minimal realization of a framework with multiple dark sectors
and provides a transparent example for illustrating the underlying dynamics.
As a result, the dominant annihilation channels of the thermal relic proceed predominantly into the darker concealed sector rather than into the SM.
The depletion of dark matter is thus governed by interactions among dark sectors,
rather than by the ultraweak portal to the SM.

\emph{Annihilating to the darker} is a general mechanism that can be naturally extended
to scenarios with additional dark sectors and to other thermal dark matter mass ranges,
where it may either help deplete the dark matter abundance or enrich the dark sector dynamics, thereby leading to a richer phenomenology.
Within this framework, two distinct possibilities naturally emerge:
\begin{itemize}
  \item \textbf{Assisted depletion:} the hidden sector thermal relic remains the dominant dark matter component today,
  but its relic abundance is efficiently depleted with the assistance of the darker sector.\footnote{
  Although our setup shares some similarities with~\cite{Belanger:2011ww}, the underlying mechanism differs in both motivation and technical treatment. \cite{Belanger:2011ww} was developed before dark matter and collider searches reached their current precision, and was primarily motivated by supplementing an \emph{insufficient} relic abundance with an additional dark matter component.
  In the present experimental situation, the primary issue is instead the \emph{overproduction} of thermal relics,
  and the main task is therefore to deplete the excess abundance efficiently,
  which cannot be achieved within~\cite{Belanger:2011ww}.
  Moreover, once the portal interactions between the extra $U(1)$ sector and the SM are forced to be ultraweak by experimental constraints,
  the assumption that the dark gauge boson remains in thermal equilibrium with the SM sector is generally no longer valid.
  In that regime, an accurate relic density calculation requires solving the full set of coupled Boltzmann equations,
  and the results of~\cite{Belanger:2011ww} are not directly applicable under current experimental constraints.}
  \item \textbf{Darker conversion:} hidden sector states primarily transfer their abundance into darker states,
  and thus the dominant dark matter in the present Universe resides in the darker sector rather than in the original hidden one.
\end{itemize}
Both cases illustrate that multi-sector dynamics can play an essential role in determining the final dark matter relic abundance,
even when the interaction connecting the hidden sector to the SM is ultraweak.

To determine the full evolution of the dark sector, it is essential to solve the coupled Boltzmann equations for all relevant dark particles, since the dark sector dynamics can be highly nontrivial. Recent work has also emphasized that, once the portal between the dark sector and the SM becomes sufficiently weak, a full coupled Boltzmann treatment is necessary for an accurate relic density calculation~\cite{Duan:2024urq}. In particular, when the dark sector interacts only ultraweakly with the SM, one cannot in general assume that the dark mediators remain in thermal equilibrium with the SM bath~\cite{Feng:2025cmi}.

The remainder of this paper is organized as follows.
Section~\ref{Sec:Portal} summarizes the general classes of portal interactions that may connect the SM,
a hidden sector, and a darker concealed sector, together with the corresponding experimental constraints.
We then consider two representative two-$U(1)$ realizations to illustrate the \emph{annihilating to the darker} mechanism.
The first is a generic $U(1)_x \times U(1)_c$ setup, which serves as a prototype framework for hidden--concealed annihilation and conversion,
and is discussed in Section~\ref{Sec:2U1}.
The second, discussed in Section~\ref{Sec:B-L}, replaces the $U(1)_x$ gauge symmetry with $U(1)_{B-L}$,
thereby providing a particularly well-motivated example in which the portal to the SM is strongly constrained
and therefore cannot by itself reproduce the observed dark matter relic abundance through secluded freeze-out,
while the darker sector may still govern the cosmological evolution of the thermal relic.
The coupled Boltzmann equations for all relevant dark particles, which determine the complete dark sector evolution,
are summarized in the Appendix. Our conclusions are given in Section~\ref{sec:Con}.

\section{Portal interactions between visible and dark sectors}\label{Sec:Portal}

We illustrate the dynamics among multiple dark sectors using a framework consisting of the SM, a hidden sector, and a concealed sector,
as shown in Fig.~\ref{Fig:Model}.
This construction is sufficiently general to apply to SM extensions containing multiple dark sectors with different structures.
Portals between the SM and the hidden sector determine how dark matter may be produced, searched for, or constrained experimentally,
and are therefore often required by experimental bounds to be ultraweak.
By contrast, portals between the hidden and concealed sectors are much less directly constrained
and may remain sizable enough to dominate the relic density depletion or dark sector conversion processes in the early Universe.
As a result, the interactions that primarily govern the cosmological evolution of dark matter
need not coincide with the portal interactions connecting the dark sector to the SM.
This separation makes it possible for thermal relic dark matter to couple only ultraweakly to the SM,
while still annihilating efficiently into a darker sector or converting into a more secluded dark relic.

\subsection{Portal interactions}

At the renormalizable level, the most common portals may be classified into scalar, gauge, and fermion portals.
In addition, axion-like particle (ALP) portals provide an important non-minimal possibility.
Below we summarize these portal interactions in a model-independent manner.

\paragraph*{Scalar portal}
A scalar portal between the SM and the hidden sector is generated by
\begin{equation}
\mathcal{L}_{\rm scalar}^{\rm SM\text{-}hid}\supset \lambda_{Hx}\,(H^\dagger H)(\Phi_x^\dagger \Phi_x)\, ,
\end{equation}
where $H$ is the SM Higgs doublet and $\Phi_x$ is a hidden sector scalar.
If $\Phi_x$ acquires a vacuum expectation value (vev), the dark Higgs scalar mixes with the SM Higgs boson, thereby modifying Higgs couplings and opening the possibility of exotic Higgs decays or additional scalar resonances.
If instead $\Phi_x$ has zero vev, the portal does not induce tree-level scalar mixing, and mainly contributes through Higgs decays into hidden scalars when kinematically allowed.
Thus the scalar portal to the SM is typically most relevant for Higgs phenomenology.

A scalar portal between the hidden and concealed sectors is described by
\begin{equation}
\mathcal{L}_{\rm scalar}^{\rm hid\text{-}con}\supset \lambda_{xc}\,(\Phi_x^\dagger \Phi_x)(\Phi_c^\dagger \Phi_c)\,,
\end{equation}
where $\Phi_c$ is a concealed sector scalar.
If both scalars develop vevs, the portal induces dark scalar mixing;
otherwise it contributes mainly through quartic interactions and dark sector scattering channels.
Unlike the SM--hidden scalar portal, this interaction does not directly involve SM fields and is therefore usually much less constrained by experiments.
Instead, it mainly affects dark sector masses, annihilation channels, and abundance transfer between the hidden and concealed sectors.

Given the chiral nature of the SM fermion spectrum, direct scalar couplings to SM fermions are tightly constrained by gauge invariance.
In contrast, vectorlike fermions can be incorporated naturally in dark sectors.
It is therefore natural to consider a real scalar $\phi$ that couples directly to vectorlike fermions from different dark sectors,
\begin{equation}
\mathcal{L}_{\rm scalar (f)}^{\rm hid\text{-}con}\supset y_x \phi \,\overline{\chi_x}\chi_x + y_c \phi \,\overline{\chi_c}\chi_c\,.
\end{equation}

Another useful possibility is provided by an ALP, which may serve as a mediator rather than as the dark matter itself.
A generic ALP $a$ may couple both to SM fields and to dark sector fields through derivative and dimension-five interactions, \emph{e.g.},
\begin{equation}
\mathcal{L}_{\rm ALP}\supset
\frac{c_i}{f_a}\,a\,F_i\widetilde{F}_i
+\frac{\partial_\mu a}{f_a}\,\overline{\psi}\gamma^\mu\gamma_5\psi\,,
\end{equation}
where $F_i$ denotes the relevant field strength, and $f_a$ is the ALP decay constant.
If the ALP couples to SM gauge bosons or fermions, it provides another portal between the SM and the hidden sector.
If it couples mainly to hidden and concealed states, it can mediate annihilation or conversion processes entirely within the dark sectors.
A generic ALP has model-dependent mass and couplings, and may lie anywhere from very light scales to the GeV--TeV regime.
Thus the ALP portal provides a particularly flexible framework for multi-sector dark dynamics.

\paragraph*{Gauge portal}
For abelian gauge symmetries, the most generic portal is kinetic mixing~\cite{Holdom:1985ag}.
Between the SM and the hidden sector, one has
\begin{equation}
\mathcal{L}_{\rm gauge}^{\rm SM\text{-}hid}\supset
-\frac{\delta_{xY}}{2}\,F_x F_{Y}\, ,
\end{equation}
where $F_x$ is the hidden $U(1)_x$ field strength and $F_Y$ is the hypercharge field strength.
After diagonalizing the kinetic and mass terms, the hidden gauge boson acquires suppressed couplings to SM currents.
This portal is often of primary phenomenological importance, as it determines collider signatures, direct detection observables,
and the decays of unstable hidden sector states, and is thus tightly constrained by experiments.

Between the hidden and concealed sectors, one has
\begin{equation}
\mathcal{L}_{\rm gauge}^{\rm hid\text{-}con}\supset
-\frac{\delta_{xc}}{2}\,F_x F_{c}
-\frac{1}{2}M_m^2\, XC\, ,
\end{equation}
where the first term represents kinetic mixing and the second term denotes mass mixing between the hidden and concealed gauge bosons~\cite{Cheung:2007ut,Feldman:2007wj,Feng:2023ubl}.
Mass mixing is common among multiple hidden sectors and can be generated through either the dark Higgs or Stueckelberg mechanism,
characterized by the parameter $M_m$.
In particular, mass mixing can have a well-motivated stringy origin~\cite{Feng:2014cla}, see~\cite{Feng:2014eja} for a general discussion.
After diagonalization, particles from one dark sector may interact with the gauge bosons of the other.
This naturally induces channels such as
\begin{equation}
\chi_x\overline{\chi_x}\to Z_c^\prime Z_c^\prime\,,\qquad
\chi_x\overline{\chi_x}\to \chi_c\overline{\chi_c}\,,
\end{equation}
or the corresponding inverse processes.
Since the $U(1)_c$ sector does not directly couple to the SM,
the hidden--concealed gauge interaction is generally much less constrained than
the direct gauge portal to the SM,
while being more important for the internal cosmological dynamics of the dark sectors.

\paragraph*{Fermion portal}
Fermion portals arise when the communication between sectors is carried directly by fermionic degrees of freedom.
Compared with scalar and gauge portals, they are usually more model-dependent, since gauge invariance constrains the allowed charge assignments and interactions.

A well-studied example of the SM--hidden fermion portal is the right-handed neutrino portal,
\begin{equation}
\mathcal{L}_{\rm fermion}^{\rm RHN}\supset
y_{\alpha I}\,\overline{L_\alpha}\,\widetilde{H}\,N_I
+y_{xI}\,\overline{\chi_x}\,\Phi_x\,N_I +h.c.\,,
\end{equation}
where the singlet fermion $N_I$ couples both to the SM and to hidden sector fields.
This portal is particularly attractive because the active--sterile mixing may be small,
while the hidden sector coupling can still be relevant for freeze-in, freeze-out, or dark sector conversion.

Another possibility is to introduce an additional connector fermion $F$ carrying SM gauge charges.
At the minimal level, one may write interactions of the form
\begin{equation}
\mathcal{L}_{\rm fermion}^{\rm charged} \supset
- y_1 \overline{F} \widetilde{H} P_R \chi_{x}
- y_2 \overline{\chi_{x}} \widetilde{H}^\dagger P_R F  - y_L \phi_x \overline{F} P_L L_i
+h.c.\,,
\end{equation}
where $F$ is an $SU(2)_L$ doublet $\sim (1,2,-\tfrac{1}{2},q_x)$,
and the dark matter candidates can be a dark fermion $\chi_{x}\sim (1,1,0,q_x)$
or a complex dark scalar $\phi_x \sim (1,1,0,-q_x)$.
Such interactions provide a direct fermionic bridge between the SM and the hidden sector,
while the additional fermions must be introduced in a way that preserves gauge anomaly cancellation.

As for hidden--concealed fermion portal,
since conventional model building usually introduces vectorlike fermions instead of chiral fermions in dark sectors,
a purely fermionic connection is difficult to realize in a minimal way by adding a separate vectorlike mediator alone.
One possibility is that some dark fermions carry charges under both $U(1)_x$ and $U(1)_c$,
thereby generating a connection through either Yukawa couplings or mass mixing terms.

\subsection{General experimental constraints}

The portal interactions introduced in the previous section are subject to qualitatively different classes of experimental constraints.
A useful general pattern is that portals directly connecting the SM to a hidden sector are often subject to significant laboratory, astrophysical, and cosmological bounds, whereas portals connecting the hidden and concealed sectors are usually much less constrained.

For scalar portals, the dominant constraints arise when the hidden scalar mixes
with the SM Higgs boson or when the SM Higgs can decay into hidden scalars.
In the former case, Higgs signal strengths and direct searches for additional scalar resonances constrain the mixing angle and the scalar mass.
In the latter case, invisible or exotic Higgs decays constrain the corresponding quartic interaction.
By contrast, scalar interactions purely between the hidden and concealed sectors do not directly modify SM Higgs couplings,
and are therefore generally much less constrained experimentally.
For ALP portals, the ALP couplings to the SM are constrained by collider searches, low-energy probes,
astrophysical limits, and cosmological observations, depending on the ALP mass and lifetime.
If the ALP couples mainly within the dark sectors, however, these bounds can be substantially relaxed.
Thus, an ALP may serve as a weak portal to the SM,
while the ALP interactions within the hidden and concealed sectors can remain substantially stronger.

For gauge portals, the most important constraints arise from the induced couplings of the dark gauge boson to SM currents. Depending on the mass range, collider searches for dilepton or dijet resonances, displaced decays, electroweak precision measurements, beam-dump experiments, and direct detection limits can all impose strong bounds on the kinetic or mass mixing that connects the hidden sector to the SM. In many viable regions of parameter space, these bounds force the SM--hidden gauge portal to be rather weak. This in turn implies that annihilation directly into SM particles, or even secluded annihilation through the dark gauge boson associated with the SM portal, may be insufficient to deplete an electroweak scale thermal relic efficiently. By contrast, gauge mixing between the hidden and concealed sectors is typically much less constrained, and may remain large enough to dominate the dark sector dynamics.

For fermion portals, the main experimental constraints depend on the nature of the mediator.
If the portal involves heavy vectorlike fermions or chiral fermions mixing with SM states,
the extended fermion sector must be chosen such that gauge anomaly cancellation is preserved.
Collider searches and precision electroweak data then place important bounds on the mixing parameters and mediator masses.

To summarize, experimental data often constrain the portal between the SM and a hidden sector to be tiny or even ultraweak,
especially when that portal would otherwise induce collider signals or dark matter scattering off SM particles.
This does not imply that thermal relic dark matter must be cosmologically ineffective.
Rather, it motivates the possibility that the interaction responsible for experimental accessibility is distinct from the interaction responsible for relic density depletion. In particular, once a darker concealed sector is present, hidden--concealed interactions may remain sizable and may efficiently drive processes such as $\chi_x\overline{\chi_x} \to Z_c' Z_c'$ and $\chi_x\overline{\chi_x} \to \chi_c\overline{\chi_c}$,
or related channels, even when direct annihilation into SM particles is ultraweak.

In the following sections, we illustrate this general picture in two simple classes of $U(1)$ models.
The first serves as a generic prototype for multi-sector annihilation and conversion,
while the second focuses on the particularly motivated case in which the direct portal to the SM is identified with $U(1)_{B-L}$.

\section{$U(1)_x\times U(1)_c$ prototype framework}\label{Sec:2U1}

To illustrate the general mechanism discussed above, we first consider a simple prototype framework
based on two abelian dark gauge symmetries $U(1)_x \times U(1)_c$,
where $U(1)_x$ denotes the hidden sector directly connected to the SM,
while $U(1)_c$ denotes a darker concealed sector.
The purpose of this setup is to provide a minimal and transparent realization of
how an ultraweak portal to the SM can still accommodate viable thermal relic dark matter
in the presence of efficient dark sector annihilation or conversion.
For simplicity, we do not include dark Higgs fields in our analysis and instead assume
that the $U(1)$ gauge bosons acquire their masses through the Stueckelberg mechanism.
Including dark Higgs fields would make the dark sector dynamics more involved,
but would also lead to a richer phenomenology, such as gravitational waves generated by $U(1)$ phase transitions~\cite{Feng:2024pab,Feng:2025wvc,Feng:2026iqo}.

\begin{figure}[t!]
\begin{center}
\includegraphics[scale=0.4]{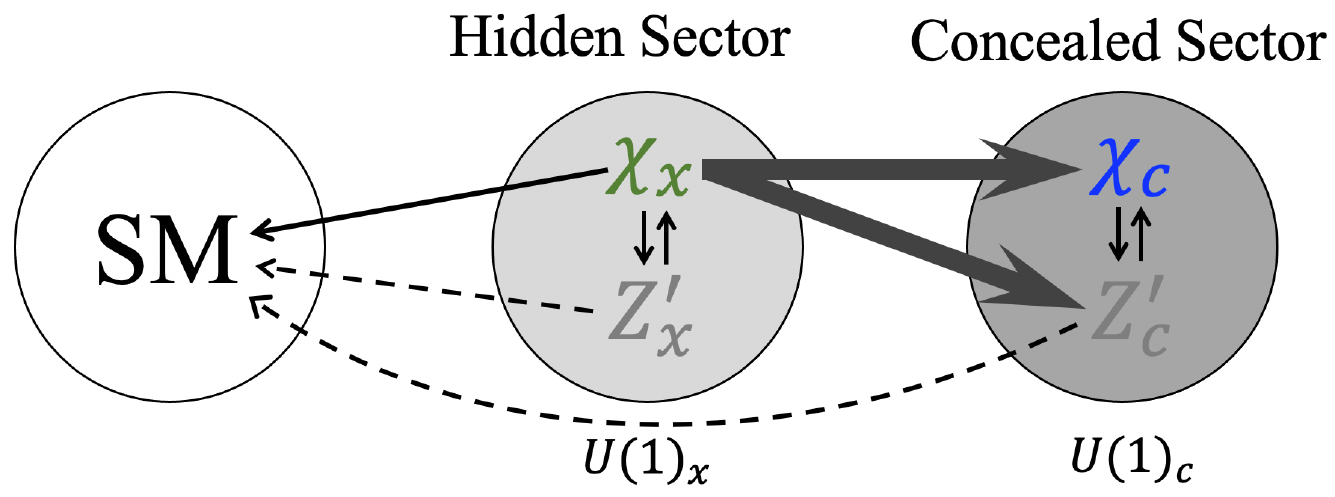}
\caption{A schematic illustration of the $U(1)_x\times U(1)_c$ framework.
The dark matter candidate $\chi_x$ predominantly freezes out into $\chi_c$ and $Z_c^\prime$ in the darker concealed $U(1)_c$ sector,
while only a small fraction annihilates into SM particles.
In the \emph{assisted depletion} scenario, $\chi_x$ remains the dominant dark matter component with $\Omega_{\chi_x} h^2 \simeq 0.12$,
while $\chi_c$ is subdominant.
In the \emph{darker conversion} scenario, $\chi_x$ is mainly converted into $\chi_c$,
with $\Omega_{\chi_c} h^2 \simeq 0.12$.}
\label{Fig:Model2U1}
\end{center}
\end{figure}

We introduce a Dirac fermion $\chi_x$ charged under $U(1)_x$ and a Dirac fermion $\chi_c$ charged under $U(1)_c$,
together with the corresponding gauge bosons $X_\mu$ and $C_\mu$.
The dark matter masses $m_{\chi_x}$ and $m_{\chi_c}$ are taken to lie at the electroweak scale, \emph{i.e.}, in the range $1-200$~GeV.
The hidden sector dark matter $\chi_x$ freezes out predominantly into the concealed $U(1)_c$ sector,
since the $U(1)_x$ sector couples more strongly to $U(1)_c$, as illustrated in Fig.~\ref{Fig:Model2U1}.
The effective Lagrangian may be written as
\begin{equation}
\mathcal{L}=\mathcal{L}_{\rm SM}+\mathcal{L}_{\rm hid}+\mathcal{L}_{\rm con}+\mathcal{L}_{\rm mix}\, ,
\end{equation}
with
\begin{align}
\mathcal{L}_{\rm hid} &\supset
-\frac14 F_x^2
- g_x X_\mu \overline{\chi_x}\gamma^\mu \chi_x
-\frac12 M_x^2 X^2 \, ,\\
\mathcal{L}_{\rm con} &\supset
-\frac14 F_c^2
- g_c C_\mu \overline{\chi_c}\gamma^\mu \chi_c
-\frac12 M_c^2 C^2 \, ,\\
\mathcal{L}_{\rm mix} &=
-\frac{\delta_{1}}{2}F_x F_Y
-\frac{\delta_{2}}{2}F_x F_c
-\frac12 M_m^2 XC \,, \label{eq:U1m}
\end{align}
Here $F_Y$, $F_x$, and $F_c$ denote the field strengths of the hypercharge, $U(1)_x$, and $U(1)_c$ gauge fields, respectively.
The parameter $\delta_1$ controls the kinetic mixing between the hidden $U(1)_x$ sector and the hypercharge gauge field,
while $\delta_2$ and $M_m$ characterize the hidden--concealed portal through kinetic and mass mixing, respectively.
We assume that SM particles are not charged under $U(1)_x$, and thus the $U(1)_x$ sector communicates with the SM only through mixing effects.

In this case, the gauge coupling $g_x$ can be sizable,
while the mixing parameter $\delta_1$ connecting the hidden sector to the SM must remain small in order to satisfy experimental constraints.
By contrast, the hidden--concealed portal is much less directly constrained and may therefore be substantially larger.
In the benchmark scenarios considered below, we take $\delta_2 \gg \delta_1$.
The $U(1)_x$ gauge boson $X_\mu$ is taken to be heavier than the dark matter particle $\chi_x$,
otherwise the dominant annihilation channel of $\chi_x$ would be into a pair of $U(1)_x$ gauge bosons~\cite{Pospelov:2007mp}.

After the kinetic and mass terms are diagonalized, the physical gauge bosons contain admixtures of both sectors.
Consequently, the hidden-sector fermion $\chi_x$ may annihilate not only into SM states,
but also more efficiently into concealed sector particles through channels such as
\begin{equation}
\chi_x \overline{\chi_x} \to \chi_c \overline{\chi_c}\,,\quad
\chi_x \overline{\chi_x} \to Z_c' Z_c'\,,\quad
\chi_x \overline{\chi_x} \to Z_x' Z_c'\,,
\end{equation}
together with the corresponding inverse processes when kinematically allowed.
Since direct annihilation into SM fermions is suppressed by the ultraweak portal to the SM,
these hidden--concealed processes can dominate the dark matter evolution.
This prototype framework naturally realizes the two possibilities previously discussed:
\begin{itemize}
  \item \emph{Assisted depletion.}
The hidden sector fermion $\chi_x$ remains the dominant dark matter component today,
but its relic density is efficiently reduced with the assistance of the concealed sector.
In this regime, the concealed sector provides the dominant additional annihilation channels,
which are not controlled by the ultraweak portal to the SM.
The hidden--concealed interactions thus control the freeze-out dynamics,
while the portal to the SM mainly determines how weakly dark matter couples to SM particles.
  \item \emph{Darker conversion.}
The hidden sector fermion $\chi_x$ is not the dominant relic in the present Universe.
Rather, it predominantly transfers its abundance into the concealed sector, and thus the final dark matter abundance is mostly carried by $\chi_c$.
This regime is particularly interesting in that the dark matter component residing in the sector directly connected to the SM
is no longer the dominant relic today.
The relic density history is then governed not simply by the hidden sector freeze-out, but by genuine multi-sector conversion dynamics.
\end{itemize}

The two cases above share the same central feature:
the interaction connecting the dark sector to the SM
need not be the same as the interaction that governs the cosmological evolution of the relic abundance.
The former is controlled by the small parameter $\delta_{1}$ and is therefore restricted
by direct detection data as well as other experimental constraints,
while the latter is governed by the hidden--concealed portal through $\delta_{2}$ and the mass mixing $M_m$.
This separation allows a thermal relic in the hidden sector to remain only ultraweakly coupled to the SM,
while still experiencing efficient depletion or conversion in the early Universe.

\section{$U(1)_{B-L}\times U(1)_c$ with a real scalar mediator}\label{Sec:B-L}

We now consider a more motivated realization in which the direct portal to the SM is provided by the gauged $U(1)_{B-L}$ symmetry,
while the connection to a darker concealed sector is mediated primarily by a real scalar field $\phi$.
This setup is particularly clean conceptually.
The $U(1)_{B-L}$ gauge boson couples directly to SM fermions through the $B-L$ current,
and thus interactions connecting the dark sector to the SM are determined by the gauge structure.
The real scalar mediator $\phi$ is assumed to couple only to hidden sector and concealed sector fermions.
We further take $\phi$ to be heavier than the concealed sector fermions, and thus it can decay completely into dark matter prior to BBN.
Since the SM fermions are chiral under $SU(2)_L\times U(1)_Y$,
a real singlet scalar cannot couple to them through a renormalizable Yukawa interaction.
Therefore, in the absence of Higgs mixing, the scalar portal to the SM is absent,
and the hidden--concealed connection may be written in a particularly simple form.

\begin{figure}[h!]
	\begin{center}
		\includegraphics[scale=0.4]{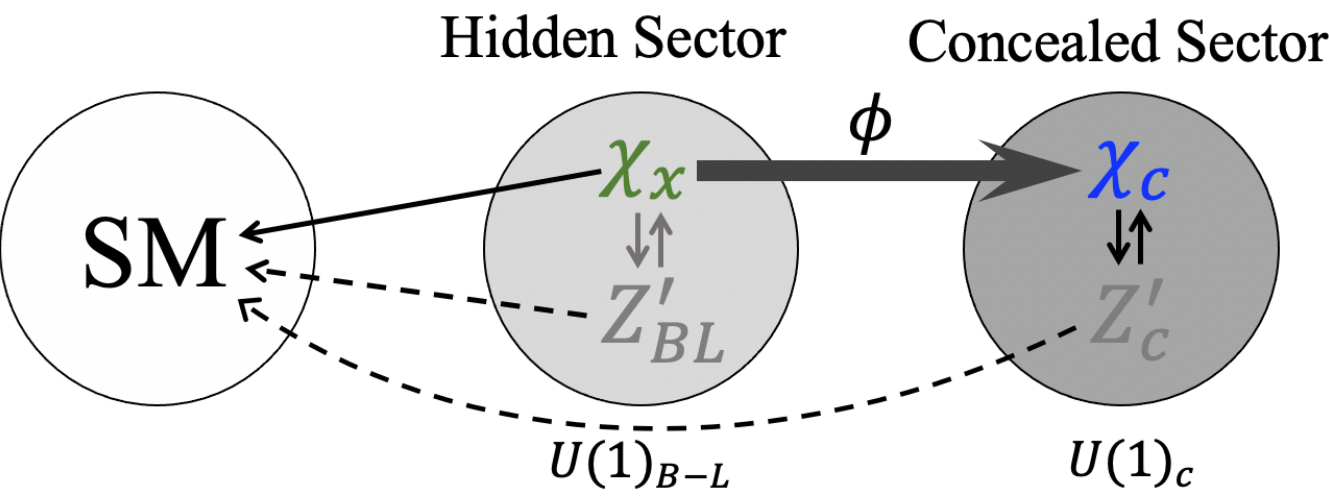}
\caption{A schematic illustration of the $U(1)_{B-L}\times U(1)_c$ framework.
The dark matter candidate $\chi_x$ is charged under $U(1)_{B-L}$ and couples to the SM through the $B-L$ gauge interaction,
while its interaction with the darker concealed $U(1)_c$ sector is mediated primarily by the real scalar $\phi$.
Since the $B-L$ gauge coupling is strongly constrained,
annihilation of $\chi_x$ directly into SM particles or into $Z_{BL}^\prime Z_{BL}^\prime$ is highly suppressed.
Instead, $\chi_x$ predominantly freezes out through hidden--concealed processes such as
$\chi_x\overline{\chi_x}\to \phi\phi$ and $\chi_x\overline{\chi_x}\to \chi_c\overline{\chi_c}$.
In the \emph{assisted depletion} scenario, $\chi_x$ remains the dominant dark matter component with $\Omega_{\chi_x}h^2\simeq 0.12$, while $\chi_c$ is subdominant.
In the \emph{darker conversion} scenario, $\chi_x$ is mainly converted into $\chi_c$, with $\Omega_{\chi_c}h^2\simeq 0.12$.}
		\label{Fig:ModelBL}
	\end{center}
\end{figure}

We introduce a Dirac fermion $\chi_x$ charged under $U(1)_{B-L}$,
a concealed $U(1)_c$ sector with gauge boson $C_\mu$, a Dirac fermion $\chi_c$, and a real scalar mediator $\phi$.
The effective Lagrangian is written as
\begin{equation}
\mathcal{L}
=
\mathcal{L}_{\rm SM}
+\mathcal{L}_{BL}
+\mathcal{L}_{\rm con}
+\mathcal{L}_{\rm portal}\, ,
\end{equation}
with
\begin{align}
\mathcal{L}_{BL}
&\supset
-\frac14 F_{BL}^2
- g_{BL} X J_{BL}
 -\frac12 M_{{BL}}^2 X^2 \, ,\\
\mathcal{L}_{\rm con} &\supset
-\frac14 F_{c}^2
- g_c C_\mu \overline{\chi_c}\gamma^\mu \chi_c
-\frac12 M_c^2 C^2 \, ,\\
\mathcal{L}_{\rm portal}
&\supset
-\frac{\delta_{2}}{2}F_{BL} F_c
- y_x \phi \overline{\chi_x} \chi_x
- y_c \phi \overline{\chi_c} \chi_c \,,
\end{align}
where $F_{BL}, F_c$ are the field strengths of $U(1)_{B-L}$, $U(1)_c$ gauge fields, respectively.
Here $J_{BL}$ denotes the $B-L$ current, and $X$ denotes the $U(1)_{B-L}$ gauge boson in the original gauge eigenbasis.
After mixing, the corresponding mass eigenstate will be denoted by $Z_{BL}^{\prime}$, see Appendix~\ref{App-B}.
For simplicity, we again do not include dark Higgs fields in our analysis,
and they can be incorporated straightforwardly and treated consistently by extending the Boltzmann equations.
We also neglect Higgs portal interactions involving $\phi$, such as $\phi H^\dagger H$,
and thus the scalar mediator does not mix with the SM Higgs boson and does not couple directly to SM fields.

The interaction of $\chi_x$ with the SM is therefore controlled entirely by the $U(1)_{B-L}$ gauge coupling $g_{BL}$.
Experimental constraints typically require the $B-L$ portal to be ultraweak for $Z_{BL}^{\prime}$ masses in the MeV--TeV range.
Hence, the annihilation of $\chi_x\overline{\chi_x}$ directly into SM particles is strongly suppressed.
The secluded channel $\chi_x\overline{\chi_x}\to Z_{BL}^{\prime}Z_{BL}^{\prime}$ is also suppressed by the ultraweak gauge coupling $g_{BL}$,
and is thus insufficient to deplete an electroweak scale thermal relic efficiently.
The role of the concealed sector is to provide additional annihilation or conversion channels which are not controlled by the ultraweak $U(1)_{B-L}$  gauge coupling.

In the present setup, the real scalar mediator opens the hidden--concealed processes
\begin{equation}
\chi_x \overline{\chi_x} \to \phi\phi\,,\qquad
\chi_x \overline{\chi_x} \to \chi_c\overline{\chi_c} \,,
\end{equation}
whenever they are kinematically allowed.
The first process proceeds through the Yukawa coupling $y_x$ and can efficiently deplete the abundance of $\chi_x$ even when the $B-L$ interactions are ultraweak.
The second process, controlled by both $y_x$ and $y_c$, transfers the hidden sector abundance into the concealed fermion $\chi_c$.
Thus the scalar-mediated hidden--concealed portal can realize the two possibilities:
\begin{itemize}
  \item \emph{Assisted depletion.}
  If $\chi_c$ is heavier than $\chi_x$, or if the direct conversion channels from $\chi_x$ to $\chi_c$ are subdominant,
  the main effect of the scalar portal is to enhance the depletion of $\chi_x$ through the channel $\chi_x\overline{\chi_x}\to \phi\phi$.
  In this case, $\chi_x$ remains the dominant dark matter component today,
  but its relic abundance is determined mainly by the hidden--concealed interaction rather than by the $U(1)_{B-L}$ interactions.
  \item \emph{Darker conversion.}
  If $\chi_c$ is lighter than $\chi_x$ and the channel $\chi_x \overline{\chi_x} \to \chi_c\overline{\chi_c}$ is efficient, the hidden sector abundance may be transferred predominantly into the concealed sector. The final dark matter component is then dominated by $\chi_c$, while the $(B-L)$-charged state $\chi_x$ survives only as a subdominant remnant.
  This provides a simple realization of dark sector conversion,
  in which the state directly connected to the SM is not the dominant relic in the present Universe.
\end{itemize}

This setup provides a novel mechanism for efficiently depleting $(B-L)$-charged dark matter, allowing electroweak scale $(B-L)$-charged thermal relics to remain viable dark matter candidates. The two roles of the dark sector interactions are clearly separated: the $U(1)_{B-L}$ gauge interaction controls the visibility of the $(B-L)$-charged dark matter and is therefore tightly constrained, while the scalar-mediated hidden--concealed interaction controls the cosmological evolution and may remain sizable. Since the mediator $\phi$ does not couple directly to SM fermions at the renormalizable level, the portal structure is especially clean:
the couplings $y_x$ and $y_c$ are not directly constrained by experimental bounds on the $U(1)_{B-L}$ gauge sector.
As a result, they may naturally dominate the annihilation or conversion processes relevant for the dark matter relic abundance.

\section{Phenomenology}\label{Sec:Pheno}

\subsection{Experimental constraints}\label{sec:Constraints}

In this work, we focus on dark matter candidates with masses in the range 1--200~GeV,
a region strongly constrained by direct dark matter searches.
Dark matter models with candidates in this mass range are also constrained by complementary indirect searches
and by other experimental probes of the dark sector, including collider and low-energy searches for the dark mediators.

\paragraph{Direct detection constraints}
The spin-independent cross section for dark matter scattering off a nucleon is given by
\begin{equation}
\sigma_{p}^{\mathrm{SI}}=\frac{\mu_{p}^{2}}{\pi}\frac{\left[Z \mathfrak{f}_{p}+\left(A-Z\right)\mathfrak{f}_{n}\right]^{2}}{A^{2}}\,,
\end{equation}
where $A,Z$ are the mass number and atomic number of the target nucleus respectively, $\mu_p=m_{p}m_{\chi}/(m_{p}+m_{\chi})$
is the dark matter--proton reduced mass.
The effective dark matter couplings to protons and neutrons $\mathfrak{f}_{p},\mathfrak{f}_{n}$
are given by the quark effective couplings
\begin{equation}
\mathfrak{f}_{p}=2\mathfrak{f}_{u}+\mathfrak{f}_{d}\,,\qquad\mathfrak{f}_{n}=\mathfrak{f}_{u}+2\mathfrak{f}_{d}\,,
\end{equation}
and are calculated from the effective operators
\begin{equation}
\mathcal{L}_{\mathrm{eff}}^u=\mathfrak{f}_{u}
\overline{\chi}\gamma_{\mu}\chi\overline{u}\gamma^{\mu}u\,,\qquad
\mathcal{L}_{\mathrm{eff}}^d=\mathfrak{f}_{d}
\overline{\chi}\gamma_{\mu}\chi\overline{d}\gamma^{\mu}d\,,
\label{eq:Leff}
\end{equation}
where the effective couplings are given by
\begin{equation}
\mathfrak{f}_{u}= \sum_{a} \frac{g_{\chi a}(g_{u_{L}a}+g_{u_{R}a})}{2M_{a}^{2}}\,,\qquad
\mathfrak{f}_{d}= \sum_{a} \frac{g_{\chi a}(g_{d_{L}a}+g_{d_{R}a})}{2M_{a}^{2}}\,.
\end{equation}
In the above expressions,
$g_{i a}$ represents the coupling of gauge bosons $a=Z,Z^\prime_x (Z_{BL}^{\prime}), Z^\prime_c$
to fermion and anti-fermion pairs $i\overline i$ where $i= \chi_x, \chi_c, u, d$.
Both $\chi_x$ from $U(1)_x/U(1)_{B-L}$ sector and $\chi_c$ from $U(1)_c$ sector are dark matter candidates and are thus subject to the
direct detection constraints.

For the dark matter mass range considered in this work,
direct detection experiments~\cite{XENON:2019zpr,XENON:2019gfn,DarkSide-50:2022qzh,PandaX-4T:2021bab,LZ:2022lsv,XENON:2023cxc,LZ:2024zvo}
place stringent constraints.
For the $U(1)_x\times U(1)_c$ setup, the spin-independent cross section is controlled by the effective coupling $\delta_{1} g_x$.
Thus, a sizable $g_x$, which can induce strong hidden--concealed sector interactions, requires a tiny kinetic mixing parameter $\delta_1$.
In the $U(1)_{B-L}\times U(1)_c$ realization, the $U(1)_{B-L}$ gauge boson couples to both dark matter and SM quarks with the same gauge coupling $g_{BL}$.
Hence, direct constraints on the $U(1)_{B-L}$ gauge boson are comparable to, or even stronger than,
the corresponding dark matter direct detection constraints.
In all benchmark models, both $\chi_x$ from the $U(1)_x$ sector and $\chi_c$ from the $U(1)_c$ sector
are verified to satisfy the direct detection constraints.

\paragraph{Indirect detection constraints}
Indirect detection searches constrain the annihilation of stable dark matter into SM particles in the present Universe.
In the analysis we used constraints from CMB~\cite{Planck:2015fie},
AMS02~\cite{AMS:2021nhj},
{\it Fermi}~\cite{McDaniel:2023bju}
and CTA~\cite{CTA:2020qlo} for various SM final states.
Dark matter annihilating into dark sectors is also restricted by
$\chi\overline{\chi}\to Z^\prime Z^\prime \to {\rm SM}$ processes for $U(1)$ models~\cite{Elor:2015bho}.
For the dark matter mass range considered in this work,
indirect detection constraints are generally less restrictive than direct detection constraints
and mediator constraints from collider searches and low-energy experiments.

\paragraph{Collider and low-energy laboratory constraints}
For $100~\textrm{MeV}-100~\textrm{GeV}$ mediators, the most relevant constraints are provided by dimuon searches at LHCb~\cite{LHCb:2019vmc} and CMS~\cite{CMS:2019buh,CMS:2023hwl}, BaBar searches for dark photons~\cite{BaBar:2014zli}, and the recent ATLAS dimuon resonance search~\cite{ATLAS:2026bpi}.
In the sub-GeV region, complementary limits are obtained from fixed-target, beam-dump, and scattering experiments.
In particular, NA62 searches in beam-dump mode constrain visibly decaying light vector mediators through leptonic final states, and provide sensitivity in the few-hundred-MeV mass range~\cite{NA62:2023nhs,NA62:2023qyn}.
Scattering measurements, including Borexino, TEXONO, GEMMA, CHARM-II~\cite{Bilmis:2015lja}, and COHERENT~\cite{AtzoriCorona:2022moj},
further constrain a light $U(1)_{B-L}$ gauge boson through deviations in neutrino--electron or coherent elastic neutrino--nucleus scattering.
A broader summary and recasting limits in a general $U(1)$ framework can be found in~\cite{Asai:2023mzl}, where $x_H=0$ corresponds to the $U(1)_{B-L}$ scenario.

\subsection{Benchmark analysis}\label{Sec:Bench}

\paragraph{The $U(1)_x \times U(1)_c$ setup}
After carefully solving the coupled Boltzmann equations given in Appendix~\ref{App-A} to track the evolution of all dark sector particles,
we present benchmark models for the $U(1)_x \times U(1)_c$ setup in Table~\ref{Tab:2U1}.
The coupling parameters are chosen to be consistent with the experimental constraints discussed in Section~\ref{sec:Constraints}.

\begin{center}
		\begin{table}[h]
            \footnotesize
			\setlength{\tabcolsep}{1.55mm}
			\begin{tabular}{|c|c|c|c|c|c|c|c|c|c|c|c|c|c|}
				\hline
				\multicolumn{1}{|c|}{Model} &
				\multicolumn{1}{c|}{$m_x$} &
				\multicolumn{1}{c|}{$m_c$} &
				\multicolumn{1}{c|}{$M_x$} &
				\multicolumn{1}{c|}{$M_c$} &
				\multicolumn{1}{c|}{$M_m$} &
				\multicolumn{1}{c|}{$M_{Z_x^{\prime}}$} &
				\multicolumn{1}{c|}{$M_{Z_c^{\prime}}$} &
				\multicolumn{1}{c|}{$\delta_{1}$} &
				\multicolumn{1}{c|}{$\delta_{2}$} &
				\multicolumn{1}{c|}{$g_{x}$} &
				\multicolumn{1}{c|}{$g_{c}$} &
				\multicolumn{1}{c|}{$\Omega_{\chi_x}h^{2}$} &
				\multicolumn{1}{c|}{$\Omega_{\chi_c}h^{2}$} \\ \hline
				
				\rowcolor{blue!10}
				\multicolumn{1}{|c|}{$a1$}
				& 200 & 90 & 210 & 80 & 100 & 211 & 65 & $1\times10^{-4}$ & 0.15 & 0.525 & 0.6 & 0.12 & $7.5\times10^{-3}$   \\ \hline		
				\rowcolor{blue!10}
				\multicolumn{1}{|c|}{$b1$}
				& 100 & 70 & 120 & 60 & 60 & 121 & 51.5 & $3\times10^{-5}$ & 0.1 & 0.4 & 0.6 & 0.12 & $2.0\times10^{-3}$  \\ \hline	
				\rowcolor{blue!10}
				\multicolumn{1}{|c|}{$c1$}
				& 10 & 6 & 15 & 4 & 3 & 15 & 4 & $6\times10^{-6}$ & 0.05 & 0.14 & 0.6 & 0.12 & $7.7\times10^{-4}$   \\ \hline	

				\rowcolor{green!10}
				\multicolumn{1}{|c|}{$a2$}
				& 200 & 165 & 210 & 160 & 160 & 221 & 103 & $4\times10^{-5}$ & 0.3 & 0.6 & 0.31 & 0.017 & 0.103   \\ \hline	
				\rowcolor{green!10}
				\multicolumn{1}{|c|}{$b2$}
				& 100 & 90 & 110 & 80 & 70 & 112 & 67 & $2\times10^{-5}$ & 0.2 & 0.6 & 0.235 & 0.010 & 0.11  \\ \hline
				\rowcolor{green!10}
				\multicolumn{1}{|c|}{$c2$}
				& 10 & 7 & 12 & 6.5 & 5 & 12 & 6.3 & $1\times10^{-6}$ & 0.2 & 0.6 & 0.075 & $1.5\times10^{-3}$ & 0.12  \\ \hline
				\rowcolor{green!10}
				\multicolumn{1}{|c|}{$c2^\prime$}
				& 10 & 7 & 12 & 9 & 5.8 & 12.6 & 8.2 & $1\times10^{-6}$ & $ 0.005 $ & 0.6 & 0.6 & $8.5\times10^{-4}$ & 0.12  \\ \hline
				
			\end{tabular}
            \caption{[All masses are in GeV.] Benchmark models for the $U(1)_x \times U(1)_c$ framework. The hidden sector dark matter candidate $\chi_x$ primarily freezes out into the concealed sector, while only a small fraction annihilates into SM states and remains undetected. Models $a1$, $b1$, and $c1$ illustrate the \emph{assisted depletion} scenario, in which $\chi_x$ attains the observed relic abundance with the assistance of the $U(1)_c$ sector. Models $a2$, $b2$, and $c2$ illustrate the \emph{darker conversion} scenario, in which $\chi_x$ primarily freezes out into the concealed sector and is converted into $\chi_c$, which becomes the dominant dark matter component. Model $c2^\prime$ represents a special case in which $\chi_c$ is the lightest particle in the entire dark sector.}
			\label{Tab:2U1}
		\end{table}
	\end{center}

Models $a1$, $b1$, and $c1$ correspond to the \emph{assisted depletion} scenario,
in which $\chi_x$ attains the observed relic abundance with the assistance of the $U(1)_c$ sector.
Although $\chi_x$ couples only ultraweakly to SM particles,
it can still annihilate efficiently through its interactions with the $U(1)_c$ sector.
For the hidden--concealed portal, we consider both kinetic and mass mixing between $U(1)_x$ and $U(1)_c$,
characterized by $\delta_2$ and $M_m$, respectively.
We emphasize that pure kinetic mixing, with $M_m=0$, is already sufficient to realize all of these cases.

Models $a2$, $b2$, $c2$ and $c2^\prime$ correspond to the \emph{darker conversion} scenario,
in which $\chi_x$ primarily freezes out into the concealed sector and is converted into $\chi_c$,
which becomes the dominant dark matter component, leaving only a small remnant of $\chi_x$ in the present Universe.
Model $c2^\prime$ represents a specific type of models
where the dark matter $\chi_c$ is the lightest among the entire hidden sector.
This differs from the usual secluded dark matter scenario~\cite{Pospelov:2007mp},
in which the mediator is typically lighter than the dark matter.

\begin{figure}[h]
\begin{center}
\includegraphics[scale=0.32]{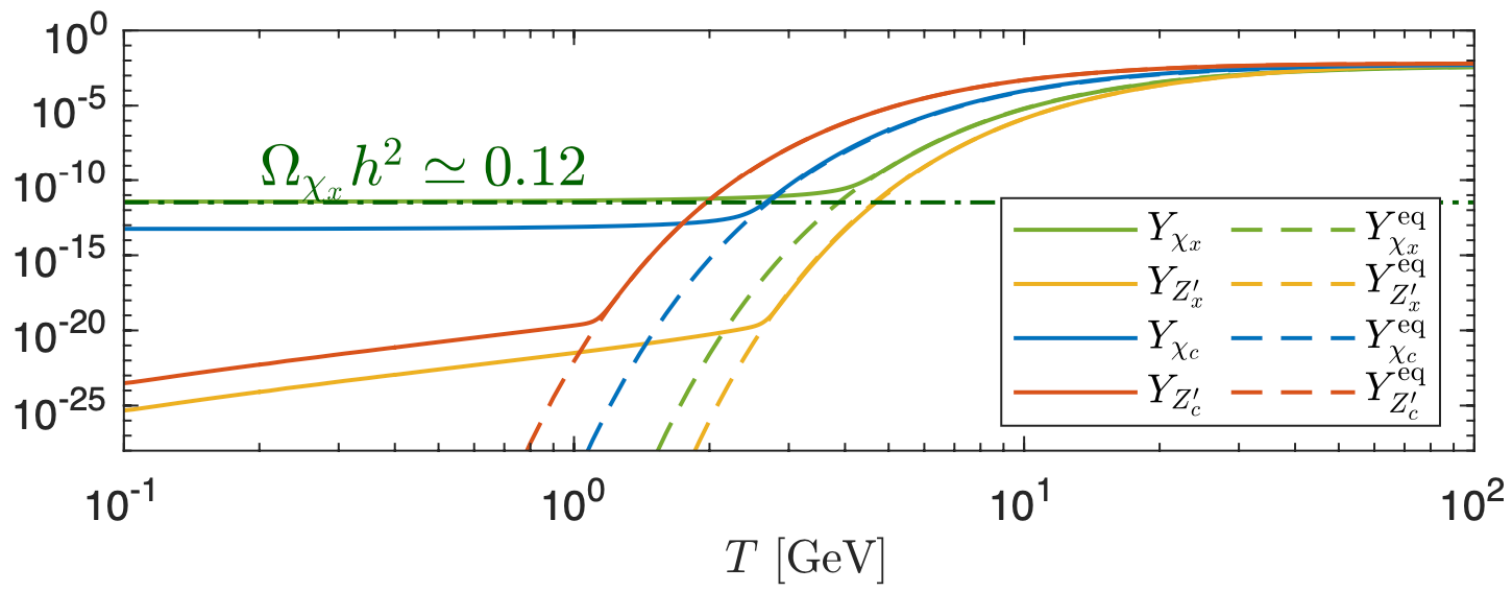}
\includegraphics[scale=0.32]{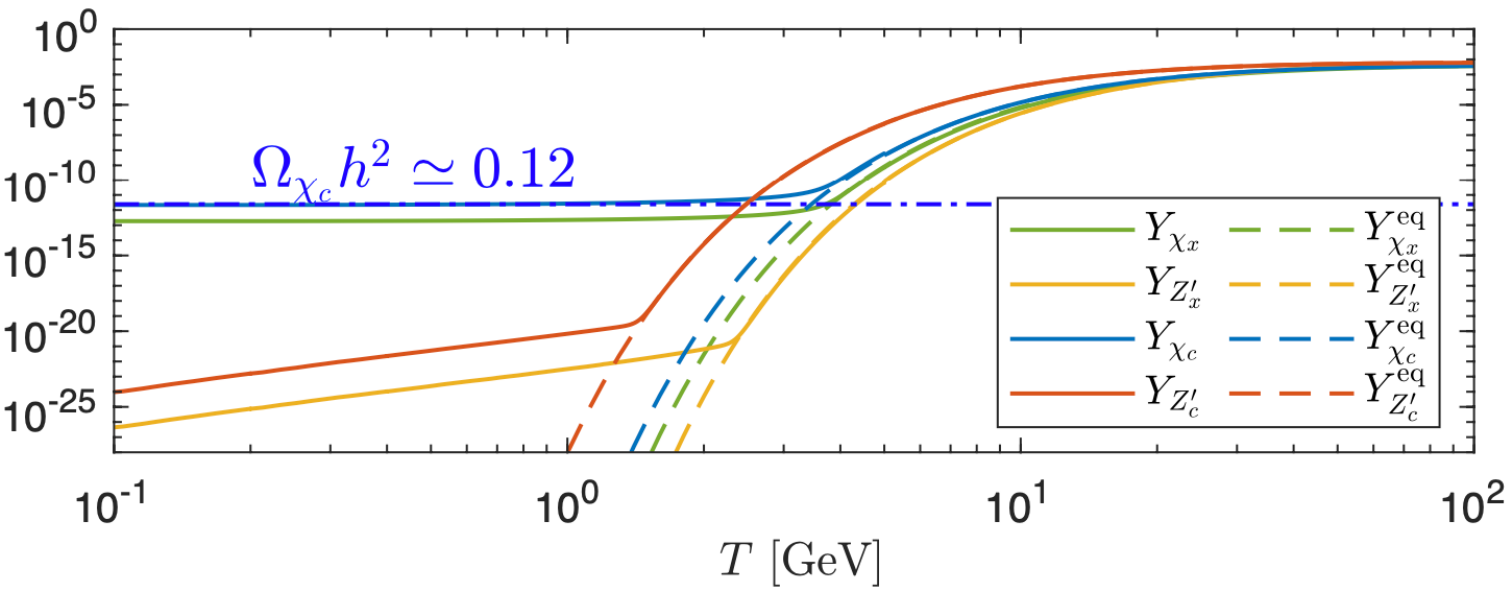}
\caption{Evolution of dark sector particles in Models $b1$ (left) and $b2$ (right),
both featuring an electroweak scale 100~GeV thermal relic $\chi_x$ that predominantly annihilates into the concealed sector.
In Model $b1$, $\chi_x$ is the dominant dark matter candidate,
whereas in Model $b2$, $\chi_c$ constitutes the dominant dark matter component.}
\label{Fig:Evo}
\end{center}
\end{figure}

Fig.~\ref{Fig:Evo} illustrates the evolution of dark particles for Models $b1$ and $b2$,
both featuring a 100~GeV thermal relic $\chi_x$.
In Model $b1$, $\chi_x$ depletes efficiently and serves as the primary dark matter candidate,
while in Model $b2$, $\chi_c$ becomes the dominant dark matter candidate.

\paragraph{The $U(1)_{B-L} \times U(1)_c$ setup}

The coupling structure and the corresponding Boltzmann equations for this setup are given in Appendix~\ref{App-B},
and representative benchmark models are shown in Table~\ref{Tab:BL}.
The parameters are chosen to satisfy all current experimental constraints discussed in Section~\ref{sec:Constraints}.
Since the $U(1)_{B-L}$ gauge coupling is strongly constrained, the annihilation channels $\chi_x\overline{\chi_x}\to f_i\overline{f_i}$ and $\chi_x\overline{\chi_x}\to Z_{BL}^{\prime}Z_{BL}^{\prime}$ are both highly suppressed, making it \emph{impossible} for a single $U(1)_{B-L}$ sector to reproduce the observed relic density with $(B-L)$-charged dark matter in the $1-200$~GeV mass range.

In the absence of Higgs mixing, the Yukawa couplings $y_x$ and $y_c$ are not directly constrained
by searches involving SM final states and can therefore remain sizable.
The relic abundance of $\chi_x$ is thus dominated by the hidden--concealed processes.
In this setup, the abundance of $\chi_x$ is efficiently transferred into the scalar mediator and concealed sector fermion through
$\chi_{x}\overline{\chi_{x}}\to\phi\phi$ and $\chi_{x}\overline{\chi_{x}}\to\chi_{c}\overline{\chi_{c}}$.
We take $m_\phi > 2m_c$, and thus the decay channel $\phi \to \chi_c \overline{\chi_c}$ is kinematically open,
thereby ensuring that $\phi$ ultimately decays into $\chi_c$.

The kinetic mixing parameter $\delta_2$, which we take to be small, also induces an indirect coupling between $Z_c^\prime$ and the SM.
Thus, $Z_c^\prime$ is constrained by collider and low-energy searches for dark mediators.
However, the effective coupling of $Z_c^\prime$ to the SM is further suppressed by the squared mass ratio, scaling as $g_{BL}\delta_2\epsilon^2/(1-\epsilon^2)$, where $\epsilon=M_c/M_{BL}$.
As a result, the experimental constraints on $Z_c^\prime$ can be readily satisfied in the benchmark models.
For small values of both $\delta_2$ and $\epsilon$, processes such as $\chi_c\overline{\chi_c}\leftrightarrow f_i\overline{f_i}$ and
$\chi_c\overline{\chi_c}\leftrightarrow Z_{BL}^{\prime}Z_{BL}^{\prime}$ are highly suppressed and remain subleading.

The final relic abundance of $\chi_c$ is further controlled by the $U(1)_c$ gauge interaction.
For sufficiently large $g_c$, the process $\chi_c\overline{\chi_c}\to Z_c^{\prime}Z_c^{\prime}$ efficiently depletes the concealed sector component,
while the produced $Z_c^{\prime}$ bosons decay into SM fermion pairs before BBN, realizing the \emph{assisted depletion} scenario.
This case is illustrated by Models $d1$, $e1$, $f1$, and $f1^\prime$, in which the $(B-L)$-charged fermion $\chi_x$ remains the dominant dark matter component and attains the observed relic abundance with the assistance of the $U(1)_c$ sector,
while $\chi_c$ contributes only a subdominant relic density.
Model $f1^\prime$ holds special interest since
asymmetric dark matter models typically predict an $\mathcal{O}(1)$~GeV dark matter particle
to explain the cosmic coincidence puzzle~\cite{Kaplan:2009ag,Feng:2012jn,Feng:2013wn,Feng:2013zda,Feng:2025wvc}.
To ensure that the asymmetric component constitutes the dominant fraction of dark matter,
the symmetric component must be depleted to below $1\%$ of the total relic density.
Within the \emph{assisted depletion} framework,
Model $f1^\prime$, with 1~GeV dark matter, can efficiently reduce the symmetric component to the $1\%$ level or below.
These benchmark models demonstrate that even when the $U(1)_{B-L}$ interaction is too weak to deplete a thermal relic on its own,
sufficiently strong hidden--concealed interactions can dominate the freeze-out of $\chi_x$ and reduce its relic abundance to the observed value.

For smaller values of $g_c$, the concealed fermion $\chi_c$ survives as the dominant relic, realizing the \emph{darker conversion} scenario.
This case is illustrated by Models $d2$, $e2$, $f2$, and $f2^\prime$, in which $\chi_x$ is converted into $\chi_c$ during freeze-out, leaving $\chi_c$ as the dominant dark matter component, while the $(B-L)$-charged fermion $\chi_x$ remains only a subdominant component in the present Universe.
Model $f2^\prime$ represents a particular class of scenarios in which the dark matter candidate $\chi_c$ is the lightest particle in the entire dark sector.

\begin{center}
	\begin{table}[h]
         \footnotesize
		\setlength{\tabcolsep}{1.85mm}
		\begin{tabular}{|c|c|c|c|c|c|c|c|c|c|c|c|}
			\hline
			\multicolumn{1}{|c|}{Model} &
			\multicolumn{1}{c|}{$m_x$} &
			\multicolumn{1}{c|}{$m_c$} &
			\multicolumn{1}{c|}{$m_\phi$} &
			\multicolumn{1}{c|}{$M_{Z_{BL}^{\prime}}$} &
			\multicolumn{1}{c|}{$M_{Z_c^{\prime}}$} &
			\multicolumn{1}{c|}{$g_{BL}$} &
			\multicolumn{1}{c|}{$g_{c}$} &
			\multicolumn{1}{c|}{$y_{x}$} &
			\multicolumn{1}{c|}{$y_{c}$} &						
			\multicolumn{1}{c|}{$\Omega_{\chi_x}h^{2}$} &
			\multicolumn{1}{c|}{$\Omega_{\chi_c}h^{2}$} \\ \hline
						
			\rowcolor{blue!10}			
			\multicolumn{1}{|c|}{$d1$}
			& 100 & 40 & 90 & 70 & 20 & $1\times10^{-4}$ & 0.3 & 0.36 & $1\times10^{-3}$ & 0.12 & $7.89\times10^{-3}$ \\ \hline	
			\rowcolor{blue!10}			
			\multicolumn{1}{|c|}{$e1$}
			& 10 & 4 & 9 & 7 & 2 & $1\times10^{-5}$ & 0.3 & 0.117 & $2\times10^{-5}$ & 0.12 & $4.1\times10^{-4}$ \\ \hline
			\rowcolor{blue!10}			
			\multicolumn{1}{|c|}{$f1$}
			& 1 & 0.4 & 0.9 & 0.7 & 0.2  & $1\times10^{-5}$ & 0.1 & 0.043 & $2\times10^{-5}$ & 0.12 & $5.3\times10^{-4}$ \\ \hline			
			\rowcolor{blue!10}
			\multicolumn{1}{|c|}{$f1^\prime$}
			& 1 & 0.4 & 0.9 & 0.7 & 0.2 & $1\times10^{-5}$ & 0.15 & 0.15 & $2\times10^{-5}$ & $0.12\times 1\%$ & $4.1\times10^{-5}$ \\ \hline		
			
			\rowcolor{green!10}
			\multicolumn{1}{|c|}{$d2$}
			& 100 & 40 & 90 & 70 & 20 & $1\times10^{-4}$ & 0.138 & 0.7 & $1\times10^{-3}$ & $1.02\times10^{-2}$ & 0.12 \\ \hline		
			\rowcolor{green!10}
			\multicolumn{1}{|c|}{$e2$}
			& 10 & 4 & 9 & 7 & 2 & $1\times10^{-5}$ & 0.05 & 0.35 & $5\times10^{-5}$ & $2.23\times10^{-3}$ & 0.12 \\ \hline	
			\rowcolor{green!10}	
			\multicolumn{1}{|c|}{$f2$}
			& 1 & 0.4 & 0.9 & 0.7 & 0.2 & $1\times10^{-5}$ & 0.017 & 0.1 & $2\times10^{-5}$ & $5.43\times10^{-3}$ & 0.12 \\ \hline					
			\rowcolor{green!10}
			\multicolumn{1}{|c|}{$f2^\prime$}
			& 1 & 0.35 & 0.9 & 0.7 & 0.45 & $1\times10^{-5}$ & 0.45 & 0.3 & $2\times10^{-5}$ & $8.62\times10^{-5}$ & 0.12 \\ \hline											
		\end{tabular}
        \caption{[All masses are in GeV.] Benchmark models for the $U(1)_{B-L} \times U(1)_c$ framework. The $(B-L)$-charged dark matter candidate $\chi_x$ primarily freezes out into the concealed sector, while only a small fraction annihilates into SM states and remains undetected. Models $d1$, $e1$, $f1$, and $f1^\prime$ illustrate the \emph{assisted depletion} scenario, in which the $(B-L)$-charged state $\chi_x$ remains the dominant dark matter component and attains the observed relic abundance with the assistance of the $U(1)_c$ sector.
        Model $f1^\prime$ demonstrates that scenarios with $\mathcal{O}(1)$~GeV asymmetric dark matter candidates can efficiently deplete the symmetric component to the $1\%$ level or below within the \emph{assisted depletion} framework. Here $\Omega_{\chi_x}h^2$ schematically denotes the symmetric contribution to the relic density, \emph{i.e.}, the combined abundance of $\chi_x$ and $\overline{\chi_x}$. In Model $f1^\prime$, the dominant dark matter component is the asymmetric $\chi_x$ abundance, assumed to arise from early Universe asymmetry generation to explain the cosmic coincidence puzzle.
        Models $d2$, $e2$, $f2$, and $f2^\prime$ illustrate the \emph{darker conversion} scenario, in which $\chi_x$ is converted into $\chi_c$ during freeze-out, leaving $\chi_c$ as the dominant dark matter component. Model $f2^\prime$ represents a particular class of scenarios in which the dark matter candidate $\chi_c$ is the lightest particle in the entire dark sector. For all benchmark cases, the hidden--concealed kinetic mixing parameter $\delta_2$ is fixed at 0.01.}
		\label{Tab:BL}
	\end{table}
\end{center}

Taken together, these two scenarios demonstrate that the $U(1)_{B-L}\times U(1)_c$ construction provides an intriguing framework in which a $(B-L)$-charged thermal relic may still remain a viable dark matter candidate in the 1--200~GeV mass range.

\section{Conclusion}\label{sec:Con}

Electroweak scale WIMPs have long been among the most extensively studied dark matter candidates,
and their continued non-observation in direct detection experiments and related searches
has posed a serious challenge to many conventional WIMP scenarios.
In this work, we explore the possibility that the absence of dark matter discoveries may be
due to more intricate interactions among multiple dark sectors beyond the SM.
Specifically, we investigate thermal relic dark matter that couples ultraweakly to the SM but interacts more strongly with a darker concealed sector,
into which it predominantly annihilates.
In such a framework, the interaction that controls the visibility of dark matter need not coincide with the interaction that determines its relic abundance. Once multiple dark sectors are present, portal interactions connecting dark sectors to the SM may be strongly constrained by current experiments and therefore remain ultraweak,
while hidden--concealed interactions can still be sizable enough to dominate the early Universe dynamics.
Hence,
thermal relic dark matter with an ultraweak portal to the SM can remain fully viable when the relic density depletion is controlled by a darker sector.

To make this idea explicit, we first summarized the general classes of portal interactions that may connect the SM, a hidden sector, and a darker concealed sector, including scalar, gauge, and fermion portals. The general pattern is that portals directly linking the SM to a hidden sector are often tightly constrained by laboratory, astrophysical, and cosmological data, whereas hidden--concealed interactions are typically much less restricted. This separation naturally opens the possibility that thermal relic dark matter may predominantly \emph{annihilate to the darker} sector rather than into SM particles, as illustrated in Fig.~\ref{Fig:Model}.

We then presented two representative realizations of this mechanism.
In the prototype $U(1)_x \times U(1)_c$ framework depicted in Fig.~\ref{Fig:Model2U1}, the hidden sector dark matter $\chi_x$ is connected only ultraweakly to the SM, while kinetic and mass mixing with a darker $U(1)_c$ sector induces efficient hidden--concealed annihilation and conversion channels. This setup realizes two distinct possibilities: \emph{assisted depletion}, in which $\chi_x$ remains the dominant relic but its abundance is efficiently reduced with the assistance of the concealed sector, and \emph{darker conversion}, in which $\chi_x$ is converted into a more secluded state $\chi_c$, which becomes the dominant dark matter component today.

We further considered a more motivated realization based on $U(1)_{B-L}\times U(1)_c$, with a real scalar mediator connecting the hidden and concealed sectors, as illustrated in Fig.~\ref{Fig:ModelBL}. In this case, the $U(1)_{B-L}$ gauge interaction is strongly constrained, and thus neither direct annihilation into SM particles nor secluded annihilation into $Z_{BL}^{\prime} Z_{BL}^{\prime}$ is sufficient to deplete an electroweak scale thermal relic efficiently. The hidden--concealed scalar-mediated interaction thus becomes essential. We showed that both \emph{assisted depletion} and \emph{darker conversion} can be realized consistently in this framework, and hence a $(B-L)$-charged thermal relic in the 1--200~GeV mass range may remain viable once the dynamics of multiple dark sectors is taken into account.

Our analysis is based on solving the full set of coupled Boltzmann equations governing the evolution of all relevant dark sector particles,
rather than relying on equilibrium assumptions that may no longer be valid when the portal to the SM is ultraweak.
The benchmark models presented in this work satisfy current direct detection, indirect detection, collider, and low-energy constraints, while explicitly exhibiting the two characteristic multi-sector scenarios discussed above.

In our benchmark analysis, we find several additional intriguing phenomenological features arising from the \emph{annihilating to the darker} mechanism. Models $c2^\prime$ and $f2^\prime$ both represent a special class of scenarios in which the dark matter candidate is the lightest state in the entire dark sector.
Model $f1^\prime$ illustrates another well-motivated class of scenarios, in which an $\mathcal{O}(1)$~GeV asymmetric dark matter component, produced by early Universe asymmetry generation mechanisms such as cogenesis, constitutes the dominant dark matter abundance and may account for the cosmic coincidence puzzle,
while the symmetric component is efficiently depleted to below $1\%$ of the total dark matter relic density through \emph{assisted depletion}.

To summarize, thermal relic dark matter in the 1--200~GeV mass range, where current direct detection experiments are particularly sensitive, may nevertheless remain viable if it can predominantly \emph{annihilate to the darker} concealed sector,
and continue to hold potential for future dark matter detections.
Moreover, \emph{annihilating to the darker} is a general mechanism that can be naturally extended to scenarios with additional dark sectors and to other thermal dark matter mass ranges. In such cases, it may either help deplete the dark matter abundance or enrich the dark sector dynamics, thereby leading to a broader range of phenomenological possibilities. We expect that the interplay between ultraweak portals to the SM and sizable darker sector interactions will continue to provide a useful framework for exploring viable thermal relic dark matter beyond the simplest single portal scenarios.\\~\\

\noindent\textbf{Acknowledgments:}

This work is
supported in part by the National Natural Science Foundation of China under Grant No. 11935009,
and Tianjin University Self-Innovation Fund Extreme Basic Research Project Grant No. 2025XJ21-0007.



\appendix

\section{Dynamics of multiple dark sectors and coupled Boltzmann equations}

\subsection{$U(1)_x\times U(1)_c$ prototype framework}\label{App-A}

We rewrite the Lagrangian in the gauge eigenbasis $V^{T}=(C,X,B,A^{3})$,
with the kinetic mixing matrix and mass mixing matrix given by
\begin{equation}
\mathcal{K}=\left(\begin{array}{cccc}
1 & \delta_{2} & 0 & 0\\
\delta_{2} & 1 & \delta_{1} & 0\\
0 & \delta_{1} & 1 & 0\\
0 & 0 & 0 & 1
\end{array}\right)\,,\quad
\mathbf{M}^{2}=
\left(\begin{array}{cc}
\begin{array}{cc}
M_{c}^{2} & M_m^2\\
M_m^2 & M_{x}^{2}
\end{array} & 0\\
0 & \begin{array}{cc}
\frac{1}{4}v^{2}g_{Y}^{2} & -\frac{1}{4}v^{2}g_{2}g_{Y}\\
-\frac{1}{4}v^{2}g_{2}g_{Y} & \frac{1}{4}v^{2}g_{2}^{2}
\end{array}
\end{array}\right)\,.
\end{equation}
To obtain the couplings of the physical gauge bosons with fermions,
a simultaneous diagonalization of both the kinetic and mass mixing matrices
brings the original basis into physical mass eigenbasis $E^{T}=(Z^\prime_c,Z^\prime_x,A_{\gamma},Z)$
with a $4\times 4$ rotation matrix $\mathcal{R}$ such that $V = \mathcal{R} E$.

The interactions between gauge bosons and fermions can be determined from
\begin{equation}
\mathcal{L}_{\mathrm{int}}  = - \left(g_{c}J_{c},g_{x}J_{x},g_{Y}J_{Y},g_{2}J_{3}\right) \mathcal{R} E\,,\label{eq:Lint}
\end{equation}
where $J_{Y}, J_{3}, J_x,J_c$ are the hypercharge current, the $SU(2)$ neutral current and the $U(1)_x, U(1)_c$ dark current respectively
\begin{gather}
J_{c}^{\mu}  = \overline{\chi_c}\gamma^{\mu}\chi_c\,,\
J_{x}^{\mu}  = \overline{\chi_x}\gamma^{\mu}\chi_x\,,\
J_{3}^{\mu}  =T_{i}^{3}\overline{f_i}\gamma^{\mu}P_{L}f_i\,,\\
J_{Y}^{\mu}  =Y_{i_{L}}\overline{f_{i}}\gamma^{\mu}P_{L}f_{i}+Y_{i_{R}}\overline{f_{i}}\gamma^{\mu}P_{R}f_{i}\,.
\end{gather}
The coupling of the $U(1)_x$ sector to the SM is of order $g_x\delta_1$,
and is chosen to satisfy both direct detection bounds on dark matter and experimental constraints on the mediator.
The couplings of the $U(1)_c$ sector to the SM are further suppressed due to the indirect mixing effect,
and the parameters associated with the $U(1)_c$ sector are also chosen such that all current experimental constraints are satisfied.

In addition to the conventional dark matter annihilation channels into SM fermions,
$\chi_x\overline{\chi_x}\to f_i\overline{f_i}$,\footnote{
The annihilation of $\chi_x\overline{\chi_x}$ to $Zh,W^+W^-$ final states is suppressed compared to
annihilation into SM fermion pairs, and thus we omit these contributions in the Boltzmann equations.}
which are insufficient to deplete the abundance of $\chi_x$,
new annihilation channels to the concealed sector
$\chi_x\overline{\chi_x} \to \chi_c \overline{\chi_c}, Z_x^\prime Z_c^\prime, Z_c^\prime Z_c^\prime$
can efficiently reduce the $\chi_x$ abundance down to the observed relic density value.
Gauge bosons $Z_x^\prime, Z_c^\prime$ decay into SM fermions prior to BBN.

The coupled Boltzmann equations which govern the evolution of all dark particles are given by
\begin{align}
	\frac{\mathrm{d}Y_{\chi_x}}{\mathrm{d}T}=&-\frac{s}{T\overline{H}}\sum_{i\in\mathrm{SM}}\Bigl\{\bigl[\bigl(Y_{\chi_x}^{\mathrm{eq}}\bigr)^{2}-Y_{\chi_x}^{2}\bigr]\left\langle \sigma v\right\rangle _{\chi_x\overline{\chi_x}\to f_i\overline{f_i}}+Y_{\chi_c}^{2}\left\langle \sigma v\right\rangle _{\chi_c\overline{\chi_c}\to \chi_x\overline{\chi_x}} \\
	&+Y_{Z_x^{\prime}}^{2}\left\langle \sigma v\right\rangle _{Z_x^{\prime}Z_x^{\prime}\to\chi_x\overline{\chi_x}}	+Y_{Z_{c}^{\prime}}^{2}\left\langle \sigma v\right\rangle_{Z_{c}^{\prime}Z_{c}^{\prime}\to\chi_x\overline{\chi_x}}+Y_{Z_x^{\prime}}Y_{Z_{c}^{\prime}}\left\langle \sigma v\right\rangle _{Z_x^{\prime}Z_{c}^{\prime}\to\chi_x\overline{\chi_x}}\nonumber \\
	&-Y_{\chi_x}^{2}\bigl(\left\langle \sigma v\right\rangle _{\chi_x\overline{\chi_x}\to \chi_c\overline{\chi_c}}+\left\langle \sigma v\right\rangle _{\chi_x\overline{\chi_x}\to Z_x^{\prime}Z_x^{\prime}}+\left\langle \sigma v\right\rangle _{\chi_x\overline{\chi_x}\to Z_{c}^{\prime}Z_{c}^{\prime}}+\left\langle \sigma v\right\rangle _{\chi_x\overline{\chi_x}\to Z_x^{\prime}Z_{c}^{\prime}}\bigr) \Bigr\}\,,\nonumber\\
\frac{\mathrm{d}Y_{\chi_c}}{\mathrm{d}T}=&-\frac{s}{T\overline{H}}\sum_{i\in\mathrm{SM}}\Bigl\{\bigl[\bigl(Y_{\chi_c}^{\mathrm{eq}}\bigr)^{2}-Y_{\chi_c}^{2}\bigr]\left\langle \sigma v\right\rangle _{\chi_c\overline{\chi_c}\to f_i\overline{f_i}}+Y_{\chi_x}^{2}\left\langle \sigma v\right\rangle _{\chi_x\overline{\chi_x}\to \chi_c\overline{\chi_c}} \\
	&+Y_{Z_x^{\prime}}^{2}\left\langle \sigma v\right\rangle _{Z_x^{\prime}Z_x^{\prime}\to\chi_c\overline{\chi_c}}
+Y_{Z_{c}^{\prime}}^{2}\left\langle \sigma v\right\rangle _{Z_{c}^{\prime}Z_{c}^{\prime}\to\chi_c\overline{\chi_c}}+Y_{Z_x^{\prime}}Y_{Z_{c}^{\prime}}\left\langle \sigma v\right\rangle _{Z_x^{\prime}Z_{c}^{\prime}\to\chi_c\overline{\chi_c}}\nonumber \\
	&-Y_{\chi_c}^{2}\bigl(\left\langle \sigma v\right\rangle _{\chi_c\overline{\chi_c}\to \chi_x\overline{\chi_x}}+\left\langle \sigma v\right\rangle _{\chi_c\overline{\chi_c}\to Z_x^{\prime}Z_x^{\prime}}+\left\langle \sigma v\right\rangle _{\chi_c\overline{\chi_c}\to Z_{c}^{\prime}Z_{c}^{\prime}}+\left\langle \sigma v\right\rangle _{\chi_c\overline{\chi_c}\to Z_x^{\prime}Z_{c}^{\prime}}\bigr)\nonumber \\
	&+\theta\left(M_{Z_x^{\prime}}-2 m_{c}\right)\bigl(-Y_{\chi_c}^{2}\left\langle \sigma v\right\rangle _{\chi_c\overline{\chi_c}\to Z_x^{\prime}}+\tfrac{1}{s}Y_{Z_x^{\prime}}\left\langle \Gamma\right\rangle _{Z_x^{\prime}\to\chi_c\overline{\chi_c}}\bigr)\bigr\}\,,\nonumber\\	
%
\frac{\mathrm{d}Y_{Z_x^{\prime}}}{\mathrm{d}T}=&-\frac{s}{T\overline{H}}\sum_{i\in\mathrm{SM}}\Bigl[Y_{i}^{2}\left\langle \sigma v\right\rangle _{f_i\overline{f_i}\to Z_x^{\prime}}-\frac{1}{s}Y_{Z_x^{\prime}}\left\langle \Gamma\right\rangle _{Z_x^{\prime}\to f_i\overline{f_i}}\\	 	
	&+Y_{\chi_x}^{2}\left(\left\langle \sigma v\right\rangle _{\chi_x\overline{\chi_x}\to Z_x^{\prime}Z_x^{\prime}}+\left\langle \sigma v\right\rangle _{\chi_x\overline{\chi_x}\to Z_x^{\prime}Z_{c}^{\prime}}\right)+Y_{\chi_c}^{2}\left(\left\langle \sigma v\right\rangle _{\chi_c\overline{\chi_c}\to Z_x^{\prime}Z_x^{\prime}}+\left\langle \sigma v\right\rangle _{\chi_c\overline{\chi_c}\to Z_x^{\prime}Z_{c}^{\prime}}\right) \nonumber \\
	&-Y_{Z_x^{\prime}}^{2}\bigl(\left\langle \sigma v\right\rangle _{Z_x^{\prime}Z_x^{\prime}\to\chi_x\overline{\chi_x}}+\left\langle \sigma v\right\rangle _{Z_x^{\prime}Z_x^{\prime}\to\chi_c\overline{\chi_c}}\bigr)-Y_{Z_x^{\prime}}Y_{Z_{c}^{\prime}}\bigl(\left\langle \sigma v\right\rangle _{Z_x^{\prime}Z_{c}^{\prime}\to\chi_x\overline{\chi_x}}+\left\langle \sigma v\right\rangle _{Z_x^{\prime}Z_{c}^{\prime}\to\chi_c\overline{\chi_c}}\bigr)\nonumber \\
	&+\theta\left(M_{Z_x^{\prime}}-2 m_{c}\right)\bigl(Y_{\chi_c}^{2}\left\langle \sigma v\right\rangle _{\chi_c\overline{\chi_c}\to Z_x^{\prime}}-\tfrac{1}{s}Y_{Z_x^{\prime}}\left\langle \Gamma\right\rangle _{Z_x^{\prime}\to\chi_c\overline{\chi_c}}\bigr)\Bigr]\,,\nonumber\\
	\frac{\mathrm{d}Y_{Z_{c}^{\prime}}}{\mathrm{d}T}=&-\frac{s}{T\overline{H}}\sum_{i\in\mathrm{SM}}\Bigl[Y_{i}^{2}\left\langle \sigma v\right\rangle _{f_i \overline{f_i}\to Z_{c}^{\prime}}-\frac{1}{s}Y_{Z_{c}^{\prime}}\left\langle \Gamma\right\rangle _{Z_{c}^{\prime}\to f_i\overline{f_i}}\\
	&+Y_{\chi_x}^{2}\left(\left\langle \sigma v\right\rangle _{\chi_x\overline{\chi_x}\to Z_c^{\prime}Z_c^{\prime}}+\left\langle \sigma v\right\rangle _{\chi_x\overline{\chi_x}\to Z_x^{\prime}Z_{c}^{\prime}}\right)+Y_{\chi_c}^{2}\left(\left\langle \sigma v\right\rangle _{\chi_c\overline{\chi_c}\to Z_c^{\prime}Z_c^{\prime}}+\left\langle \sigma v\right\rangle _{\chi_c\overline{\chi_c}\to Z_x^{\prime}Z_{c}^{\prime}}\right) \nonumber \\
	&-Y_{Z_{c}^{\prime}}^{2}\bigl(\left\langle \sigma v\right\rangle _{Z_{c}^{\prime}Z_{c}^{\prime}\to\chi_x\overline{\chi_x}}+\left\langle \sigma v\right\rangle _{Z_{c}^{\prime}Z_{c}^{\prime}\to\chi_c\overline{\chi_c}}\bigr)	-Y_{Z_x^{\prime}}Y_{Z_{c}^{\prime}}\bigl(\left\langle \sigma v\right\rangle _{Z_x^{\prime}Z_{c}^{\prime}\to\chi_x\overline{\chi_x}}+\left\langle \sigma v\right\rangle _{Z_x^{\prime}Z_{c}^{\prime}\to\chi_c\overline{\chi_c}}\bigr)\Bigr]\,,\nonumber
\end{align}
where $s= \frac{2\pi^{2}}{45}h_{\mathrm{eff}}T^{3}$ and
\begin{equation}
\overline{H}=  \frac{H}{1+\frac{1}{3}\frac{T}{h_{\mathrm{eff}}}\frac{\mathrm{d}h_{\mathrm{eff}}}{\mathrm{d}T}}
=\sqrt{\frac{\pi^{2}g_{\mathrm{eff}}}{90}} \frac{T^{2}/M_{\mathrm{Pl}}}{1+\frac{1}{3}\frac{T}{h_{\mathrm{eff}}}\frac{\mathrm{d}h_{\mathrm{eff}}}{\mathrm{d}T}}\,.
\end{equation}

\subsection{$U(1)_{B-L}\times U(1)_c$ with a real scalar mediator}\label{App-B}

The kinetic mixing between $U(1)_{B-L}$ and hypercharge is expected to be subdominant compared with the direct $U(1)_{B-L}$ gauge interaction under current experimental constraints, and is therefore assumed to be small and neglected in our discussion.
The electroweak sector is therefore decoupled from the following discussion, and thus the mixing between $U(1)_{B-L}$ and $U(1)_c$ reduces to a $2\times2$ system in the basis $V^T=(C,X)$, where $C$ and $X$ denote the gauge bosons of $U(1)_c$ and $U(1)_{B-L}$, respectively.

The corresponding kinetic and mass matrices are
\begin{equation}
	\mathcal{K}=\left(\begin{array}{cc}
		1 & \delta_{2} \\
		\delta_{2} & 1
	\end{array}\right)\,,\qquad
	\mathbf{M}^{2}=
		\left(\begin{array}{cc}
				M_{c}^{2} & 0\\
				0 & M_{BL}^{2}
			\end{array} \right)\,.
\end{equation}
The physical eigenstates are denoted by $E^T=(Z_c^{\prime},Z_{BL}^{\prime})$, and are related to the original gauge basis through $V=\mathcal{R}E$. After diagonalizing both kinetic and mass matrices, one obtains
\begin{gather}
	\mathcal{R}=\left(\begin{array}{cc}
		c_{\delta}\cos\theta & -c_{\delta}\sin\theta\\
		\sin\theta-s_{\delta}\cos\theta & \cos\theta+s_{\delta}\sin\theta
	\end{array}\right)\,, \label{eq:BLmix}
\end{gather}
where $c_{\delta}=1/\sqrt{1-\delta_{2}^{2}}$, $s_{\delta}=\delta_{2}/\sqrt{1-\delta_{2}^{2}}$, and the mixing angle is given by
\begin{equation}
	\theta=\frac{1}{2}\arctan\left(\frac{2\delta_{2}\sqrt{1-\delta_{2}^{2}}}{1-\epsilon^{2}-2\delta_{2}^{2}}\right)\,,
\end{equation}
with $\epsilon\equiv M_{c}/M_{BL}$. In the limit $\delta_{2}\ll1$ and $\epsilon\neq 1$, the angle reduces to $\theta\simeq\delta_{2}/(1-\epsilon^{2})$.  The physical masses of the two neutral dark gauge bosons are obtained as
\begin{equation} M_{Z_{BL}^{\prime}}=\sqrt{M_{c}^{2}\mathcal{R}_{12}^{2}+M_{BL}^{2}\mathcal{R}_{22}^{2}}\,,\qquad
M_{Z_{c}^{\prime}}=\sqrt{M_{c}^{2}\mathcal{R}_{11}^{2}+M_{BL}^{2}\mathcal{R}_{21}^{2}}\,.
\end{equation}
For $\delta_{2}\ll1$, one has $M_{Z_{BL}^{\prime}}\simeq M_{BL}$ and $M_{Z_{c}^{\prime}}\simeq M_{c}$.

The new gauge interactions in the original gauge eigenbasis are
\begin{equation}
	\mathcal{L}_{\mathrm{int}}= - g_{BL}XJ_{BL} - g_{c}CJ_{c}\,,\label{eq:Lint_BL}
\end{equation}
where
\begin{equation}
J_{BL}  =\sum_{i}Q_{i}^{BL}\overline{f_{i}}\gamma^{\mu}f_{i}+Q_{x}^{BL}\overline{\chi_{x}}\gamma^{\mu}\chi_{x}\,,\qquad J_{c}  =Q_{c}\overline{\chi_{c}}\gamma^{\mu}\chi_{c}\,,
\end{equation}
with the following charge assignments
\begin{equation}
	Q_{q}^{BL}=+\tfrac{1}{3}\,,\;Q_{\ell}^{BL}=-1\,,\;Q_{x}^{BL}=+1\,,\;Q_{c}=+1\,.
\end{equation}
Using the rotation matrix Eq.~\eqref{eq:BLmix}, gauge interactions in mass eigenbasis are written as
\begin{align}
	\mathcal{L}_{\mathrm{int}} & = - Z_{BL}^{\prime}\Bigl[\bigl(\cos\theta+s_{\delta}\sin\theta\bigr)g_{BL}J_{BL}-c_{\delta}\sin\theta g_{c}J_{c}\Bigr]\nonumber \\
	 &\quad - Z_{c}^{\prime}\Bigl[\bigl(\sin\theta-s_{\delta}\cos\theta\bigr)g_{BL}J_{BL}+c_{\delta}\cos\theta g_{c}J_{c}\Bigr]\,.
\end{align}
For $\delta_{2}\ll1$, the leading couplings can be summarized as follows:
\begin{center}
	\setlength{\tabcolsep}{8pt}
	\renewcommand{\arraystretch}{1.5}
	\begin{tabular}{l|ccc}
	\hline
	& $f_i\overline{f_i}$ & $\chi_x\overline{\chi_x}$ & $\chi_c\overline{\chi_c}$ \\
	\hline
	$Z_{BL}^{\prime}$           & $g_{BL}Q_i^{BL}$ & $g_{BL}Q_x^{BL}$ & $-g_c\frac{\delta_2}{1-\epsilon^2}$ \\
	$Z_{c}^{\prime}$            & $g_{BL}Q_i^{BL}\frac{\delta_2\epsilon^2}{1-\epsilon^2}$ & $g_{BL}Q_x^{BL}\frac{\delta_2\epsilon^2}{1-\epsilon^2}$ & $g_cQ_c$ \\
	\hline
	\end{tabular}
\end{center}


The $U(1)_{B-L}$ annihilation channels $\chi_x\overline{\chi_x}\to f_i\overline{f_i}$ and $\chi_x\overline{\chi_x}\to Z_{BL}^{\prime}Z_{BL}^{\prime}$ are highly suppressed by the ultraweak $U(1)_{B-L}$ gauge coupling $g_{BL}$ and are therefore negligible compared with the hidden--concealed processes, such as $\chi_x\overline{\chi_x}\to \phi\phi$ and $\chi_x\overline{\chi_x}\to \chi_c\overline{\chi_c}$, mediated by the real scalar $\phi$.

The full Boltzmann equations in the $U(1)_{B-L}\times U(1)_c$ setup are written as follows
\begin{align}
	\frac{\mathrm{d}Y_{\chi_{x}}}{\mathrm{d}T}= & -\frac{s}{T\overline{H}}\sum_{i\in\mathrm{SM}}\Bigl\{\bigl[\bigl(Y_{\chi_x}^{\mathrm{eq}}\bigr)^{2}-Y_{\chi_x}^{2}\bigr]\left\langle \sigma v\right\rangle _{\chi_x\overline{\chi_x}\to f_i\overline{f_i}}\nonumber \\
	&\qquad +Y_{\chi_{c}}^{2}\left\langle \sigma v\right\rangle _{\chi_{c}\overline{\chi_{c}}\to\chi_{x}\overline{\chi_{x}}}+Y_{\phi}^{2}\left\langle \sigma v\right\rangle _{\phi\phi\to\chi_{x}\overline{\chi_{x}}}+Y_{Z_{BL}^{\prime}}^{2}\left\langle \sigma v\right\rangle _{Z_{BL}^{\prime}Z_{BL}^{\prime}\to\chi_{x}\overline{\chi_{x}}}\nonumber \\
	&\qquad  -Y_{\chi_{x}}^{2}\left(\left\langle \sigma v\right\rangle _{\chi_{x}\overline{\chi_{x}}\to\chi_{c}\overline{\chi_{c}}}+\left\langle\sigma v\right\rangle _{\chi_{x}\overline{\chi_{x}}\to\phi\phi}+\left\langle\sigma v\right\rangle _{\chi_{x}\overline{\chi_{x}}\to Z_{BL}^{\prime}Z_{BL}^{\prime}}\right)\Bigr\}\,,\\
	\frac{\mathrm{d}Y_{Z_{BL}}^{\prime}}{\mathrm{d}T}= & -\frac{s}{T\overline{H}}\sum_{i\in\mathrm{SM}}\Bigl[Y_{i}^{2}\left\langle \sigma v\right\rangle _{f_{i}\overline{f_{i}}\to Z_{BL}^{\prime}}-\tfrac{1}{s}Y_{Z_{BL}^{\prime}}\left\langle \Gamma\right\rangle _{Z_{BL}^{\prime}\to f_{i}\overline{f_{i}}}+Y_{\chi_x}^{2}\left\langle\sigma v\right\rangle _{\chi_{x}\overline{\chi_{x}}\leftrightarrow Z_{BL}^{\prime}Z_{BL}^{\prime}}\nonumber \\
	&+Y_{\chi_c}^{2}\left\langle\sigma v\right\rangle _{\chi_{c}\overline{\chi_{c}}\leftrightarrow Z_{BL}^{\prime}Z_{BL}^{\prime}}-Y_{Z_{BL}^{\prime}}^{2}\left(\left\langle\sigma v\right\rangle _{Z_{BL}^{\prime}Z_{BL}^{\prime}\to\chi_{x}\overline{\chi_{x}}}+\left\langle\sigma v\right\rangle _{Z_{BL}^{\prime}Z_{BL}^{\prime}\to\chi_{c}\overline{\chi_{c}}}
	\right)\Bigr]\,,\\
%
	\frac{\mathrm{d}Y_{\chi_{c}}}{\mathrm{d}T}= & -\frac{s}{T\overline{H}}\sum_{i\in\mathrm{SM}}\Bigl\{\bigl[\bigl(Y_{\chi_c}^{\mathrm{eq}}\bigr)^{2}-Y_{\chi_c}^{2}\bigr]\left\langle \sigma v\right\rangle _{\chi_c\overline{\chi_c}\to f_i\overline{f_i}}
+Y_{\chi_{x}}^{2}\left\langle \sigma v\right\rangle _{\chi_{x}\overline{\chi_{x}}\to\chi_{c}\overline{\chi_{c}}}
\nonumber \\
	&\qquad +Y_{Z_{c}^{\prime}}^{2}\left\langle \sigma v\right\rangle _{Z_{c}^{\prime}Z_{c}^{\prime}\to\chi_{c}\overline{\chi_{c}}}+Y_{\phi}^{2}\left\langle \sigma v\right\rangle _{\phi\phi\to\chi_{c}\overline{\chi_{c}}}+Y_{Z_{BL}^{\prime}}^{2}\left\langle \sigma v\right\rangle _{Z_{BL}^{\prime}Z_{BL}^{\prime}\to\chi_{c}\overline{\chi_{c}}}\nonumber \\
	&\qquad -Y_{\chi_{c}}^{2}\left(\left\langle \sigma v\right\rangle _{\chi_{c}\overline{\chi_{c}}\to \chi_{x}\overline{\chi_{x}}}+\left\langle \sigma v\right\rangle _{\chi_{c}\overline{\chi_{c}}\to Z_{c}^{\prime}Z_{c}^{\prime}}+\left\langle \sigma v\right\rangle _{\chi_{c}\overline{\chi_{c}}\to\phi\phi}+\left\langle \sigma v\right\rangle _{\chi_{c}\overline{\chi_{c}}\to Z_{BL}^{\prime}Z_{BL}^{\prime}}\right)
	\nonumber \\
	&\qquad +\theta\bigl(m_{\phi}-2m_{\chi_{c}}\bigr)\,\bigl(-Y_{\chi_{c}}^{2}\left\langle \sigma v\right\rangle _{\chi_{c}\overline{\chi_{c}}\to\phi}+\tfrac{1}{s}Y_{\phi}\left\langle \Gamma\right\rangle _{\phi\to\chi_{c}\overline{\chi_{c}}}\bigr)\Bigr\}\,,\\
%
	\frac{\mathrm{d}Y_{Z_{c}^{\prime}}}{\mathrm{d}T}
= & -\frac{s}{T\overline{H}}\sum_{i\in\mathrm{SM}}\Bigl(Y_{i}^{2}\left\langle \sigma v\right\rangle _{f_{i}\overline{f_{i}}\to Z_{c}^{\prime}}
-\tfrac{1}{s}Y_{Z_{c}^{\prime}}\left\langle \Gamma\right\rangle _{Z_{c}^{\prime}\to f_{i}\overline{f_{i}}}\Bigr.\nonumber\\
&\qquad\qquad\quad \Bigl.
+Y_{\chi_{c}}^{2}\left\langle \sigma v\right\rangle _{\chi_{c}\overline{\chi_{c}}\to Z_{c}^{\prime}Z_{c}^{\prime}}
-Y_{Z_{c}^{\prime}}^{2}\left\langle \sigma v\right\rangle _{Z_{c}^{\prime}Z_{c}^{\prime}\to\chi_{c}\overline{\chi_{c}}}
\Bigr)\,,\\
	\frac{\mathrm{d}Y_{\phi}}{\mathrm{d}T}= & -\frac{s}{T\overline{H}}\Bigl[-Y_{\phi}^{2}\left(\left\langle \sigma v\right\rangle _{\phi\phi\to\chi_{x}\overline{\chi_{x}}}+\left\langle \sigma v\right\rangle _{\phi\phi\to\chi_{c}\overline{\chi_{c}}}\right)+Y_{\chi_{x}}^{2}\left\langle \sigma v\right\rangle _{\chi_{x}\overline{\chi_{x}}\to\phi\phi}+Y_{\chi_{c}}^{2}\left\langle \sigma v\right\rangle _{\chi_{c}\overline{\chi_{c}}\to\phi\phi}\nonumber \\
	&\qquad\quad +\theta\bigl(m_{\phi}-2m_{\chi_{c}}\bigr)\,\bigl(Y_{\chi_{c}}^{2}\left\langle \sigma v\right\rangle _{\chi_{c}\overline{\chi_{c}}\to\phi}-\tfrac{1}{s}Y_{\phi}\left\langle \Gamma\right\rangle _{\phi\to\chi_{c}\overline{\chi_{c}}}\bigr)\Bigr]\,.
\end{align}

\end{document}